\documentstyle[11pt,aaspp4]{article}
\begin{document}

\title{$BVRI$ Observations of the Optical Afterglow of GRB
990510\footnote{Based on the observations collected at the Las
Campanas Observatory 1~m Swope telescope}}

\author{K. Z. Stanek\altaffilmark{2} and P. M. Garnavich}
\affil{Harvard-Smithsonian Center for Astrophysics, 60 Garden St., MS20,
Cambridge, MA 02138}
\affil{\tt e-mail: kstanek@cfa.harvard.edu, peterg@cfa.harvard.edu}
\author{J. Kaluzny}
\affil{Warsaw University Observatory, Al. Ujazdowskie 4,
PL-00-478 Warszawa, Poland\\
and\\
N.~Copernicus Astronomical Center, 
Bartycka 18, PL-00-716 Warszawa, Poland}
\affil{\tt e-mail: jka@sirius.astrouw.edu.pl}
\author{W. Pych}
\affil{Warsaw University Observatory, Al. Ujazdowskie 4,
PL-00-478 Warszawa, Poland} 
\affil{\tt e-mail: pych@sirius.astrouw.edu.pl} 
\author{I. Thompson}
\affil{Carnegie Observatories, 813 Santa Barbara St., Pasadena, CA 91101-1292}
\affil{\tt e-mail: ian@ociw.edu}
\altaffiltext{2}{On leave from N.~Copernicus Astronomical Center, 
Bartycka 18, Warszawa PL-00-716, Poland} 

\begin{abstract}

We present $BVRI$ observations of the optical counterpart to the
Gamma-Ray Burst (GRB) 990510 obtained with the Las Campanas 1.0-m
telescope between 15 and 48 hours after the burst. The temporal
analysis of the data indicates steepening decay, independent of
wavelength, approaching asymptotically $t^{-0.76\pm 0.01}$ at early
times ($t\ll 1\;day)$ and $t^{-2.40\pm 0.02}$ at late times, with the
break time at $t_0=1.57\pm 0.03\;days$. GRB 990510 is the most rapidly
fading of the well-documented GRB afterglows. It is also the first
observed example of broad-band break for a GRB optical counterpart.
The optical spectral energy distribution, corrected for significant
Galactic reddening, is well fitted by a single power-law with
$\nu^{-0.61\pm 0.12}$. However, when the $B$-band point is dropped
from the fit, the power-law becomes $\nu^{-0.46\pm 0.08}$, indicating
a possible deviation from the power-law in the spectrum, either
intrinsic or due to additional extinction near the source or from
an intervening galaxy at $z=1.62$. Broad-band break behavior broadly
similar to that observed in GRB 990510 has been predicted in some
jet models of GRB afterglows, thus supporting the idea that the
GRB energy is beamed, at least in some cases.

\end{abstract}


\section{INTRODUCTION}

The BeppoSAX satellite (Boella et al.~1997) has brought a new
dimension to gamma-ray burst (GRB) research, by rapidly providing good
localization of several bursts per year. This has allowed many GRBs to
be followed up at other wavelengths, including X-ray (Costa et
al.~1997), optical (van Paradijs et al.~1997) and radio (Frail et
al.~1997).  Good positions have also allowed redshifts to be measured
for a number of GRBs (e.g.GRB 970508: Metzger et al.~1997), providing
a definite proof for their cosmological origin.

GRB 990510 was detected by BeppoSAX (Piro et al.~1999) on May 10.36743
UT, and the fluence was among the highest of the BeppoSAX localized
events, after GRB 990123, GRB 980329, and GRB 970111 (Amati et
al.~1999). It was also detected by BATSE (Kippen et al.~1999) and its
peak flux and fluence rank it in the top 4\% (9\%) of BATSE burst flux
(fluence) distributions.  The BeppoSAX NFIs follow-up of GRB 990510
started about 8 hours after the burst (Kuulkers et al.~1999) and
detected a strong X-ray afterglow.

Axelrod, Mould \& Schmidt (1999) were the first to optically monitor
the field of GRB 990510, starting on 10.514 UT, i.e.~only $3.5\;hours$
after the burst, using the Mount Stromlo 50-inch telescope.  The
optical counterpart to GRB 990510 was first identified by Vreeswijk et
al.~(1999a) with data taken about $8.5\; hours$ (May 10.73 UT) after
the burst, using the Sutherland 1.0-m telescope. It was easily
recognized as a bright ($R\approx19.2$; but see below), new object not
present in the DSS image at the position of $\alpha
=13^h38^m07^s.64,\;\; \delta = -80^\circ 29{'}48{''}.8\;\;{\rm
(J2000.0)}$ (Kaluzny et al. 1999b).  Galama et al.~(1999) confirmed
the optical transient (OT) with the ESO 2.2-m telescope on May 10.99
UT and found it to be declining with a temporal decay index of
$-0.85$. Kaluzny et al.~(1999a,b) observed the GRB 990510 with the Las
Campanas 1.0-m telescope, beginning on May 10.995, and found that the
object declined by $0.34\;$mag in the $R$-band over $4.6\;hours$.
They also noticed that the $R$-band reference star used by Vreeswijk
et al.~(1999) and Galama et al.~(1999) is in fact brighter than they
assumed by $0.7\;$mag (USNO-A2.0 catalog: Monet et al.~1996), making
the afterglow as bright as $R\approx 17.75$ when first observed by
Axelrod et al.~(1999). This change in calibration was confirmed by
independent photometric calibration of the GRB field by Pietrzy\'nski
\& Udalski (1999a) and Bloom et al.~(1999). Absorption lines at
$z=1.619$ seen in the optical spectrum of GRB 990510 taken with the
VLT-UT1 8-m telescope by Vreeswijk et al.~(1999b) provide a lower
limit to the redshift of the GRB source.

By adding data taken during their second night (starting May 12.0 UT),
Stanek et al.~(1999) noticed the steepening of the OT decline, and
they derived a power-law decay index of $-1.36$ by a fit to their data
from both nights, but they also showed that a single power-law does
not provide a good fit to all the data. Analysis of more data has
confirmed this behavior (Hjorth et al.~1999; Bloom et al.~1999;
Marconi et al.~1999a). We will discuss in detail the temporal
behavior of the GRB 990510 OT further in this paper.

We describe the data and the reduction procedure in Section 2.  In
Section 3 we discuss the multiband temporal behavior of the GRB OT.
In Section 4 we describe the broad-band spectral properties of the
afterglow deduced from our optical data.

\section{THE DATA AND THE REDUCTION PROCEDURE}

\begin{figure}[t]
\plotfiddle{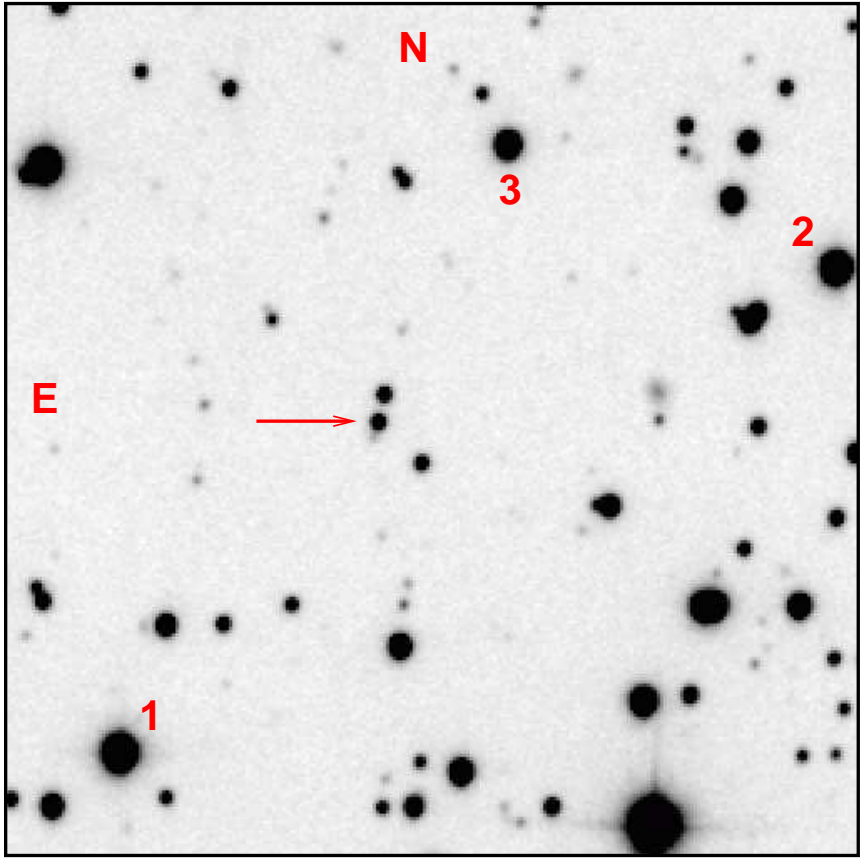}{10.5cm}{0}{70}{70}{-200}{360}
\caption{Finding chart for the field of GRB 990510. The optical
transient is indicated with an arrow. Also shown are the three
comparison stars calibrated by Pietrzy\'nski \& Udalski (1999a,b) and
Bloom et al.~(1999).}
\label{fig:field}
\end{figure}

The data were obtained with the Las Campanas Observatory 1.0-meter
Swope telescope on two nights, May 10/11 and May 11/12 UT, 1999. We
used the backside-illuminated SITE3 $2048\times 4096$ CCD, with the
pixel scale of $0.44\; arcsec\; pixel^{-1}$. To speed up the readout
time we used only a $2048\times 1200$ section of the CCD. We have
obtained 35 $R$-band, 20 $V$-band, 16 $I$-band and 4 $B$-band images,
with exposure times ranging from $240\;sec$ during the first night to
$900\;sec$ during the second night\footnote{$BVRI$ data are available
through {\tt anonymous ftp} on {\tt cfa-ftp.harvard.edu}, in {\tt
pub/kstanek/GRB990510} directory, and through the {\tt WWW} at {\tt
http://cfa-www.harvard.edu/cfa/oir/Research/GRB/}.} The image quality
was good, with median seeing for the $R$-band of $1.4\arcsec$. The
finding chart for the GRB 990510 is shown in Figure~\ref{fig:field}.

The data were reduced using two different methods. First, we have used
the photometric data pipeline of the the DIRECT project (Kaluzny et
al.~1998; Stanek et al.~1998), which is based on the Daophot
PSF-fitting image reduction package (Stetson 1987; 1992). To check the
consistency of the differential photometry we have also employed the
image subtraction method of Alard \& Lupton (1999) and Alard (1999a),
as implemented in the ISIS image subtraction package (Alard 1999b). We
have found a very good agreement between these two data reduction
methods ($\sim 0.01\;$mag).

The calibration for the field has been obtained by Pietrzy\'nski \&
Udalski (1999a,b) in the $BVI$ bands and by Bloom et al.~(1999) in the
$VR$ bands. In both cases the quoted uncertainty of the calibration
zero point is $\pm 0.03\;$mag for the $VRI$ bands and $\pm 0.05\;$mag
for the $B$-band. Comparison of the $V$-band measurements for three
secondary standards given by Pietrzy\'nski \& Udalski (1999a,b) and
Bloom et al.~(1999) in the GRB field yields a systematic difference of
$0.025\;mag$, i.e. within the quoted uncertainty. We have adopted the
calibration of Pietrzy\'nski \& Udalski (1999a,b) for the $BVI$ bands
and the calibration of Bloom et al.~(1999) for the $R$-band.

\section{THE TEMPORAL BEHAVIOR}

\begin{figure}[t]
\plotfiddle{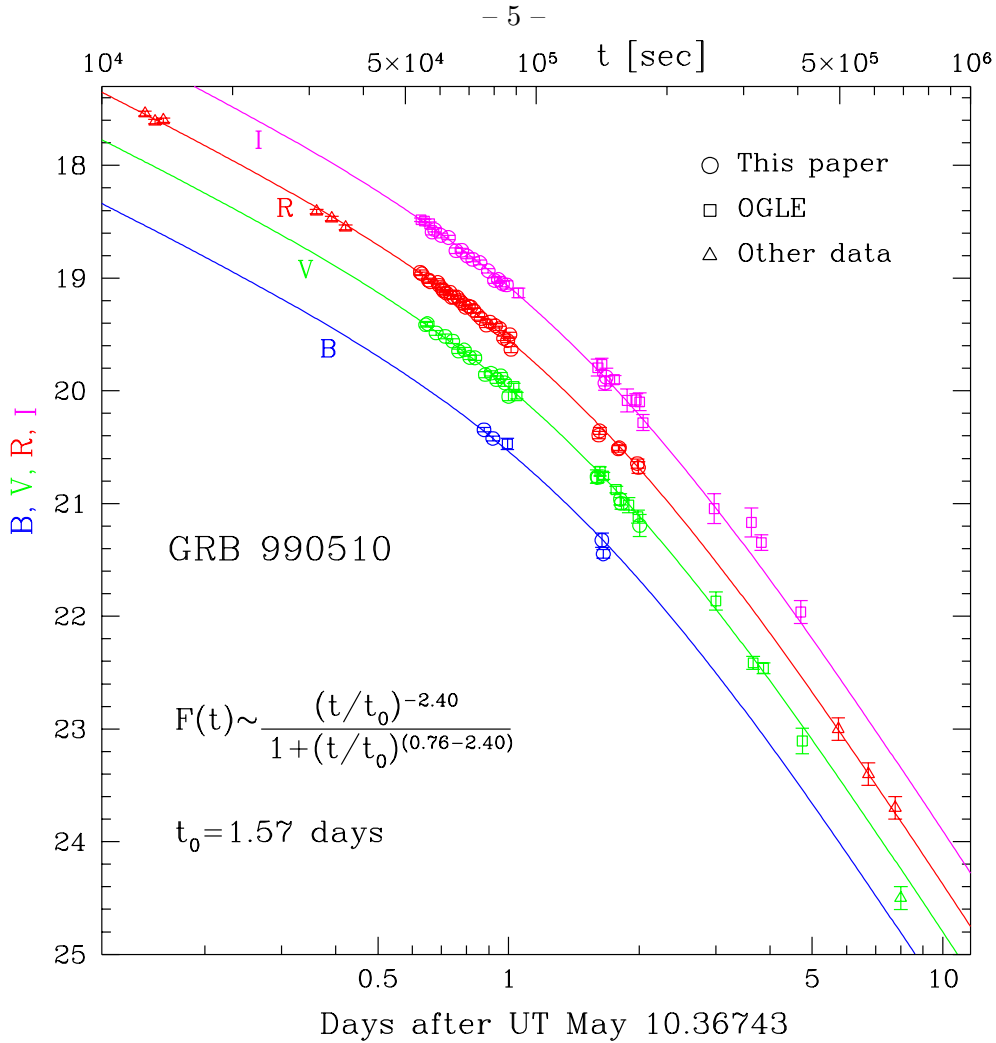}{12cm}{0}{68}{68}{-210}{-110}
\caption{$BVRI$ light curves of GRB 990510. Our data is shown with
circles and OGLE data with squares. Other data used to constrain the
fits is shown with triangles (for references see text). Also shown is
the simple analytical fit discussed in the text.}
\label{fig:time}
\end{figure}

We plot the GRB 990510 $BVRI$ lightcurves in Figure~\ref{fig:time}.
Most of the data comes from our monitoring (Kaluzny et al.~1999a,b;
Stanek et al.~1999) and from the OGLE project (Pietrzy\'nski \&
Udalski 1999a,b,c). We use those OGLE points which extend the time
coverage beyond that of our data. In addition, we use the early
$R$-band data of Axelrod et al.~(1999) and Galama et al.~(1999) and
later $R$-band data points of Marconi et al.~(1999a,b), and one late
$V$-band point obtained by Beuermann, Reinsch \& Hessman (1999). All
these data extend beyond our time coverage and are therefore useful
for constraining the time evolution of the afterglow.

As noticed by Stanek et al.~(1999) using data from two days only, the
optical $R$-band data already showed clear departure from the initial
shallow power-law of about $-0.9$ determined by Galama et al.~(1999)
and Kaluzny et al.~(1999a,b) from the first day data.  This trend of
steepening decay was confirmed later by Hjorth et al.~(1999) and
others.

To describe the temporal evolution of the GRB 990510 optical
counterpart, we use the combined data described earlier and fit to
each $VRI$ band separately the following four parameters formula
(similar to that of Bloom et al.~1999 and Marconi et al.~1999a):

\begin{equation}
F_{\nu}(t) = \frac{k_{\nu} \left(\frac{t}{t_0}\right)^{-a_1}}
{1+\left(\frac{t}{t_0}\right)^{(a_2-a_1)}},
\end{equation}

where $k_{\nu}$ is a normalization constant specific for each band
such that the magnitude is given by $m(t)=-2.5 \log(F_{\nu})$.  This
formula describes power-law $t^{-a_2}$ decline at early times ($t\ll
t_0$) and another power-law $t^{-a_1}$ decline at late times ($t\gg
t_0$). The total number of points used for the fits was 44 for the
$R$-band, 32 for the $V$-band, 31 for the $I$-band and five for the
$B$-band, total of 112 points. The $R$-band data has the most
extensive temporal coverage, especially early after the burst. The
other bands are not as well constrained, at least with data at hand,
which demonstrates the importance of getting multi-band follow-up
images as early as possible. The parameters are correlated with each
other, so we first fit the $R$-band and then use the derived $t_0$
value as fixed for the $VI$ bands. We also run a combined fit to all
$BVRI$ data. The results of this combined fit are shown as the
continuous lines in the Figure~\ref{fig:time} and the best-fit
parameters for all the fits are presented in Table~\ref{table:par}.
The errors shown are based on conditional probability distributions,
fixing two of the parameters at their most probable values.

\tablenum{1}
\begin{planotable}{ccccc}
\tablewidth{26pc}
\tablecaption{\sc GRB 990510 Best-Fit Parameters}
\tablehead{ \colhead{} & \colhead{} &
\colhead{} & \colhead{$t_0$} & \colhead{} \\  
\colhead{Band} & \colhead{$a_1$}  & \colhead{$a_2$} &
\colhead{(days)} & \colhead{$\chi^2/DOF$} }
\startdata  
$R$    & $2.41\pm 0.02$ & $0.76\pm 0.01$ & $1.55\pm 0.03$ & 1.51 \\
$V$    & $2.46\pm 0.05$ & $0.73\pm 0.02$ & \nodata 	  & 1.22 \\
$I$    & $2.17\pm 0.07$ & $0.82\pm 0.04$ & \nodata        & 1.03 \\
$BVRI$ & $2.40\pm 0.02$ & $0.76\pm 0.01$ & $1.57\pm 0.03$ & 1.48
\enddata
\label{table:par}
\end{planotable}

The combined fit does a good job of representing the overall temporal
behavior of the $BVRI$ data set, but it is clearly dominated by the
$R$-band data. Considering that these data represent an inhomogeneous
data set, the $\chi^2/DOF$ values for the $VI$ bands are rather good.
The $\chi^2/DOF$ is somewhat higher for the $R$-band. As our $R$-band
data has somewhat better photometric accuracy than the $VI$ data, this
might indicate that we see some small, short-timescales departures
from the analytical fit.  Our frequent observations of the GRB optical
counterpart during the first night were obtained to test for such
rapid variations. After we subtract the long-term trend derived above,
we get a root-mean-square ($rms$) scatter of $0.02\;$mag, with the
largest deviation of $0.08\;$mag. Comparing to constant field stars
with similar magnitude, the observed scatter is consistent with
photometric noise.  Clearly, brightness variations on time scales from
0.1 to $2\;hours$ are small.  We note that Hjorth et al.~(1999)
obtained similar $rms$ scatter in their 31 $300\;sec$ $R$-band images,
obtained with the 1.54-m telescope on La Silla at times overlapping
with our observations. By combining these two data sets one might be
able to search with more sensitivity for correlated variations in the
brightness of the afterglow.

\section{REDDENING AND BROAD-BAND SPECTRAL ENERGY DISTRIBUTION}

The GRB 990510 is located at Galactic coordinates of
$l=304\arcdeg\!\!.9426, b=-17\arcdeg\!\!.8079$. To remove the effects of
the Galactic interstellar extinction we used the reddening map of
Schlegel, Finkbeiner \& Davis (1998, hereafter: SFD). As noticed by
Stanek (1999), the expected Galactic reddening towards the burst is
significant, $E(B-V)=0.203$, which corresponds to expected values of
Galactic extinction $A_B=0.88, A_V=0.66, A_R=0.53$ and $A_I=0.40$, for
the Landolt (1992) CTIO filters and standard reddening curve of
Cardelli, Clayton \& Mathis (1989).

We synthesize the $BVRI$ spectrum from our data by interpolating the
magnitudes to a common time.  As discussed in the previous
section, the optical colors of the GRB 990510 counterpart do not show
significant variation.  We therefore select an epoch of May 11.26 UT
($21.5\;hours$ after the burst) for the color analysis, which
coincides with our first $B$-band images.  Using the all-band fit from
the previous section we get for this epoch $B_{21.5}=20.39\pm 0.05$,
$V_{21.5}=19.82\pm 0.03$, $R_{21.4}=19.40\pm 0.03$ and
$I_{21.5}=18.93\pm 0.03$, where the $0.03\;$mag error bar represents
the combined uncertainty in the zero points calibration and the small
measurement errors. Using the SFD extinction values, this translates
to unreddened values of $B_{0,21.5}=19.51\pm 0.10$,
$V_{0,21.5}=19.16\pm 0.07$, $R_{0,21.5}=18.87\pm 0.06$ and
$I_{0,21.5}=18.53\pm 0.05$ (following SFD we took the error in
$E(B-V)$ to be $0.02\;$mag).

\begin{figure}[t]
\plotfiddle{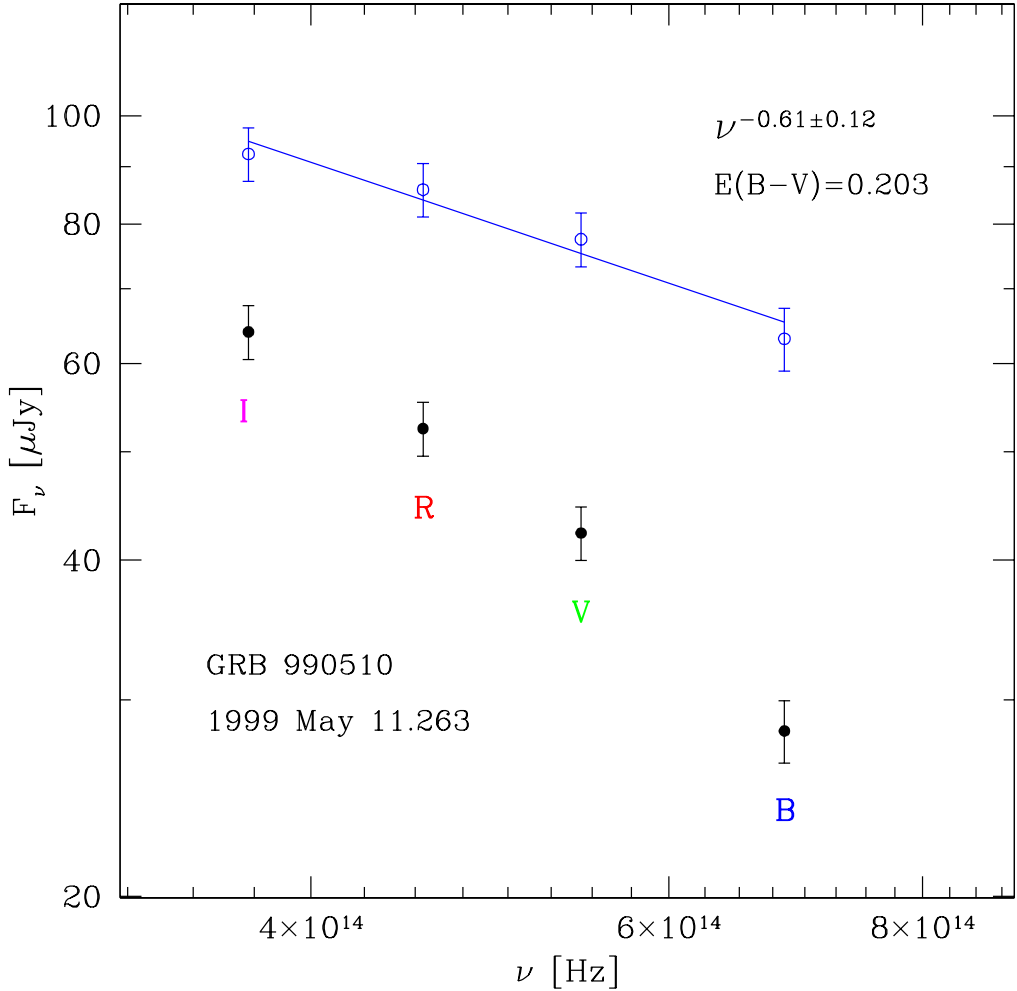}{12cm}{0}{70}{70}{-210}{-110}
\caption{Synthetic spectrum of GRB 990510 $21.5\;hours$ after the
burst, constructed using analytical fit shown in
Figure~\ref{fig:time}.}
\label{fig:spectrum}
\end{figure}

We convert the $BVRI$ magnitudes to fluxes using the effective
wavelengths and normalizations of Fukugita, Shimasaku \& Ichikawa
(1995). These conversions are accurate to about 5\%, which increases
the error-bars correspondingly. Note that while the error in the
$E(B-V)$ reddening value has not been applied to the error-bars of
individual points, we include it in the error budget of the fitted
slope. The results are plotted in Figure~\ref{fig:spectrum} for both
the observed and the dereddened magnitudes.  The spectrum is well
fitted by a single power-law with $\nu^{-0.61\pm 0.12}$, although when
the $B$-band point is dropped from the fit, the power-law becomes
$\nu^{-0.46\pm 0.08}$, indicating a possible deviation from the
power-law in the blue part of the spectrum, either intrinsic or due to
additional extinction outside our Galaxy.

\section{CONCLUSIONS}

We presented $BVRI$ observations of the GRB 990510.  Our analysis of
the data indicates steepening decay, independent of optical
wavelength, with power-law behavior $t^{-0.76\pm 0.01}$ at early times
($t\ll 1\;day)$ and second power-law $t^{-2.40\pm 0.02}$ at late
times, with the break time at $t_0=1.57\pm 0.03\; days$. This is the
first observed example of such broad-band break for a GRB OT and it is
very well documented thanks to a concerted effort of the
astronomical community. We would like to stress the importance of
multiband optical observations for the GRB studies, especially early
after the burst. 

The optical spectral energy distribution, corrected for significant
Galactic reddening, is well fitted by a single power-law with
$\nu^{-0.61\pm 0.12}$. However, when the $B$-band point is dropped
from the fit, the power-law becomes $\nu^{-0.46\pm 0.08}$, indicating
a possible deviation from the power-law in the spectrum, either
intrinsic or due to additional extinction near the host or at the
intervening galaxy at $z=1.62$.

The temporal behavior of GRB 990510 is broadly similar to predictions
of GRB jet models (Rhoads 1999).  A break in the light curve is
expected when the jet makes the transition to sideways expansion after
the relativistic Lorentz factor drops below the inverse of the opening
angle of the initial beam. After the break, the the temporal power-law
index is expected to approach the electron energy distribution index
with values between 2.0 and 2.5, which is consistent with the
late-time decline observed here for GRB 990510 (Halpern et al.~1999;
Sari, Piran \& Halpern~1999).  However, the early-time decay index and
the spectral slope of GRB~990510 are not as well explained by the
models.  GRB 990510 provides the first direct evidence that, in some
cases, GRB energy is not ejected isotropically.

\acknowledgments{S. Barthelmy, the organizer of the GRB Coordinates
Network (GCN), is recognized for his extremely useful effort. The OGLE
collaboration is thanked for making their data promptly available to
the astronomical community. We thank C. Alard for his support with the
ISIS image subtraction package. We thank B. Paczy\'nski and
L. A. Phillips for useful discussions and comments on the paper.  KZS
was supported by the Harvard-Smithsonian Center for Astrophysics
Fellowship. JK was supported by NSF grant AST-9528096 to Bohdan
Paczy\'nski and by the Polish KBN grant 2P03D011.12. WP was supported
by the Polish KBN grant 2P03D010.15}

\newpage

\tablenum{2}
\begin{planotable}{cccc}
\tablewidth{20pc}
\tablecaption{\sc GRB 990510 $B$-Band Lightcurve}
\tablehead{ \colhead{HJD} & \colhead{} & 
\colhead{} & \colhead{Hours after} \\  
\colhead{$-2450000$} & \colhead{$B$}  & 
\colhead{$\sigma_B$} & \colhead{the burst} }
\startdata  
   1309.7494 & 20.345 &  0.018  &  21.0636 \\
   1309.7914 & 20.422 &  0.021  &  22.0716 \\
   1310.5116 & 21.325 &  0.062  &  39.3564 \\
   1310.5230 & 21.446 &  0.040  &  39.6300 
\enddata
\label{table:B}
\end{planotable}

\tablenum{3}
\begin{planotable}{cccc}
\tablewidth{20pc}
\tablecaption{\sc GRB 990510 $V$-Band Lightcurve}
\tablehead{ \colhead{HJD} & \colhead{} & 
\colhead{} & \colhead{Hours after} \\  
\colhead{$-2450000$} & \colhead{$V$}  & 
\colhead{$\sigma_V$} & \colhead{the burst} }
\startdata  
   1309.5163 & 19.414  & 0.020  &  15.4692 \\
   1309.5216 & 19.404  & 0.018  &  15.5964 \\
   1309.5525 & 19.486  & 0.022  &  16.3380 \\
   1309.5857 & 19.517  & 0.020  &  17.1348 \\
   1309.6152 & 19.559  & 0.022  &  17.8428 \\
   1309.6387 & 19.648  & 0.019  &  18.4068 \\
   1309.6621 & 19.637  & 0.021  &  18.9684 \\
   1309.6852 & 19.703  & 0.020  &  19.5228 \\
   1309.7086 & 19.707  & 0.025  &  20.0844 \\
   1309.7551 & 19.856  & 0.026  &  21.2004 \\
   1309.7822 & 19.844  & 0.023  &  21.8508 \\
   1309.8096 & 19.901  & 0.034  &  22.5084 \\
   1309.8303 & 19.866  & 0.029  &  23.0052 \\
   1309.8511 & 19.929  & 0.032  &  23.5044 \\
   1309.8716 & 20.050  & 0.039  &  23.9964 \\
   1310.4680 & 20.762  & 0.059  &  38.3100 \\
   1310.4763 & 20.769  & 0.045  &  38.5092 \\
   1310.6795 & 20.962  & 0.051  &  43.3860 \\
   1310.6885 & 20.999  & 0.052  &  43.6020 \\
   1310.8748 & 21.194  & 0.099  &  48.0732
\enddata
\label{table:V}
\end{planotable}

\tablenum{4}
\begin{planotable}{cccc}
\tablewidth{20pc}
\tablecaption{\sc GRB 990510 $R$-Band Lightcurve}
\tablehead{ \colhead{HJD} & \colhead{} & 
\colhead{} & \colhead{Hours after} \\  
\colhead{$-2450000$} & \colhead{$R$}  & 
\colhead{$\sigma_R$} & \colhead{the burst} }
\startdata  
   1309.4975 & 18.949 &  0.022  &  15.0180 \\
   1309.5030 & 18.962 &  0.015  &  15.1500 \\
   1309.5255 & 19.018 &  0.018  &  15.6900 \\
   1309.5309 & 19.031 &  0.017  &  15.8196 \\
   1309.5582 & 19.038 &  0.015  &  16.4748 \\
   1309.5637 & 19.064 &  0.017  &  16.6068 \\
   1309.5756 & 19.096 &  0.015  &  16.8924 \\
   1309.5806 & 19.115 &  0.015  &  17.0124 \\
   1309.5909 & 19.133 &  0.014  &  17.2596 \\
   1309.6042 & 19.125 &  0.017  &  17.5788 \\
   1309.6094 & 19.169 &  0.015  &  17.7036 \\
   1309.6204 & 19.168 &  0.015  &  17.9676 \\
   1309.6336 & 19.170 &  0.013  &  18.2844 \\
   1309.6439 & 19.200 &  0.016  &  18.5316 \\
   1309.6570 & 19.226 &  0.021  &  18.8460 \\
   1309.6671 & 19.259 &  0.018  &  19.0884 \\
   1309.6801 & 19.250 &  0.019  &  19.4004 \\
   1309.6905 & 19.256 &  0.019  &  19.6500 \\
   1309.7035 & 19.287 &  0.021  &  19.9620 \\
   1309.7203 & 19.326 &  0.019  &  20.3652 \\
   1309.7362 & 19.357 &  0.015  &  20.7468 \\
   1309.7603 & 19.417 &  0.021  &  21.3252 \\
   1309.7770 & 19.389 &  0.022  &  21.7260 \\
   1309.8044 & 19.423 &  0.020  &  22.3836 \\
   1309.8246 & 19.447 &  0.020  &  22.8684 \\
   1309.8453 & 19.535 &  0.019  &  23.3652 \\
   1309.8659 & 19.554 &  0.020  &  23.8596 \\
   1309.8774 & 19.501 &  0.020  &  24.1356 \\
   1309.8844 & 19.633 &  0.026  &  24.3036 \\
   1310.4841 & 20.393 &  0.032  &  38.6964 \\
   1310.4923 & 20.356 &  0.032  &  38.8932 \\
   1310.6613 & 20.516 &  0.029  &  42.9492 \\
   1310.6695 & 20.505 &  0.025  &  43.1460 \\
   1310.8502 & 20.645 &  0.040  &  47.4828 \\
   1310.8618 & 20.681 &  0.052  &  47.7612
\enddata
\label{table:R}
\end{planotable}

\tablenum{5}
\begin{planotable}{cccc}
\tablewidth{20pc}
\tablecaption{\sc GRB 990510 $I$-Band Lightcurve}
\tablehead{ \colhead{HJD} & \colhead{} & 
\colhead{} & \colhead{Hours after} \\  
\colhead{$-2450000$} & \colhead{$I$}  & 
\colhead{$\sigma_I$} & \colhead{the burst} }
\startdata  
   1309.5383 & 18.592 &  0.026 &   15.9972 \\
   1309.5472 & 18.570 &  0.019 &   16.2108 \\
   1309.5705 & 18.621 &  0.021 &   16.7700 \\
   1309.5989 & 18.640 &  0.018 &   17.4516 \\
   1309.6285 & 18.757 &  0.024 &   18.1620 \\
   1309.6519 & 18.753 &  0.017 &   18.7236 \\
   1309.6750 & 18.804 &  0.024 &   19.2780 \\
   1309.6984 & 18.834 &  0.021 &   19.8396 \\
   1309.7307 & 18.862 &  0.021 &   20.6148 \\
   1309.7689 & 18.935 &  0.020 &   21.5316 \\
   1309.7994 & 19.021 &  0.025 &   22.2636 \\
   1309.8188 & 19.012 &  0.025 &   22.7292 \\
   1309.8395 & 19.051 &  0.026 &   23.2260 \\
   1309.8602 & 19.060 &  0.035 &   23.7228 \\
   1310.5375 & 19.931 &  0.064 &   39.9780 \\
   1310.5490 & 19.878 &  0.080 &   40.2540
\enddata
\label{table:I}
\end{planotable}

\end{document}